# Onset field for Fermi-surface reconstruction in the cuprate superconductor $YBa_2Cu_3O_y$

G. Grissonnanche[1], F. Laliberté[1], S. Dufour-Beauséjour[1], A. Riopel[1], S. Badoux[1],
M. Caouette-Mansour[1], M. Matusiak[1,†], A. Juneau-Fecteau[1], P. Bourgeois-Hope[1],
O. Cyr-Choinière[1], J. C. Baglo[2], B. J. Ramshaw[2,‡], R. Liang[2,3], D. A. Bonn[2,3],
W. N. Hardy[2,3], S. Krämer[4], D. LeBoeuf[4] , D. Graf[5], N. Doiron-Leyraud[1], and
Louis Taillefer[1,3]

1 *Département de physique & RQMP, Université de Sherbrooke, Sherbrooke, Québec J1K 2R1, Canada*

2 *Department of Physics & Astronomy, University of British Columbia, Vancouver, British Columbia V6T 1Z1, Canada*

3 *Canadian Institute for Advanced Research, Toronto, Ontario M5G 1Z8, Canada*

4 *Laboratoire National des Champs Magnétiques Intenses, Grenoble, France*

5 *National High Magnetic Field Laboratory, Tallahassee, Florida, USA*

---

[†] Present address : Institute of Low Temperature and Structure Research, Polish Academy of Sciences, Wroclaw 50-950, Poland.

[‡] Present address : Los Alamos National Laboratory, Los Alamos, New Mexico 87545, USA.

**Quantum oscillations[1,2] and negative Hall[3,4,5] and Seebeck[5,6,7] coefficients at low temperature and high magnetic field have shown the Fermi surface of underdoped cuprates to contain a small closed electron pocket. It is thought to result from a reconstruction by charge order, but whether it is the order seen by NMR[8,9] and ultrasound[10] above a threshold field or the short-range modulations seen by X-ray diffraction in zero field[11,12,13,14,15,16,17] is unclear. Here we use measurements of the thermal Hall conductivity in $YBa_2Cu_3O_y$ to show that Fermi-surface reconstruction occurs only above a sharply defined onset field, equal to the transition field seen in ultrasound. This reveals that electrons do not experience long-range broken translational symmetry in the zero-field ground state, and hence in zero field there is no quantum critical point for the onset of charge order as a function of doping.**

To make sense of the complex temperature-doping phase diagram of cuprate superconductors, it is essential to determine what symmetries are broken, below what temperature and at what $T = 0$ quantum critical point. The detection of charge-density-wave (CDW) order by NMR in $YBa_2Cu_3O_y$ (YBCO) at dopings near $p = 0.12$ shows that translational symmetry is broken in that cuprate below a temperature $T_{NMR}$ (Fig. 1), but only above a threshold magnetic field $H_{NMR}$ (refs. 8,9). Short-range CDW modulations are seen by X-ray diffraction (XRD) not only in YBCO (refs. 11,12), also near $p = 0.12$, but in several other cuprates as well[13,14,15], showing that CDW ordering is a generic tendency. However, those modulations are observed at temperatures well above $T_{NMR}$ (Fig. 1), and even in zero field. At what temperature is translational symmetry truly broken? The discovery of a field-induced thermodynamic phase transition in YBCO at $p = 0.11$, detected as an anomaly in sound velocities at a critical field $H_{ultrasound} = 18 \pm 0.5$ T (ref. 10), has made the situation more puzzling. While it is tempting to attribute this transition to CDW order, its onset field is inconsistent with both XRD and NMR, since $H_{NMR}$ = 10.4 ± 1.0 T at $p$ = 0.11 (ref. 9). At what field is translational symmetry broken?

In order to elucidate these puzzles, and in particular establish at what field translational symmetry is broken in YBCO ($H$ = 0, $H_{NMR}$ or $H_{ultrasound}$), we have studied its Fermi surface as a function of magnetic field. By introducing a new periodicity, CDW order causes a reduction of the Brillouin zone and a concomitant reconstruction of the Fermi surface. Therefore, a good way to ascertain whether electrons experience broken translational symmetry is to determine whether their Fermi surface is reconstructed. The signature of Fermi-surface reconstruction (FSR) is the presence of a small closed electron-like pocket, as seen in YBCO and $HgBa_2CuO_{4+\delta}$ via quantum oscillations[1,2] and negative Hall[3,4,5] and Seebeck[5,6,7] coefficients at low temperature.

FSR has so far been studied only in the normal state, *i.e.* in magnetic fields greater than the critical field $H_{c2}$ needed to suppress superconductivity[18]. Here we investigate FSR below $H_{c2}$, by measuring the thermal Hall effect, which is the transverse temperature gradient (along $y$) generated by a longitudinal thermal current (along $x$) in the presence of a perpendicular magnetic field (along $z$).

In the same way that the electrical Hall conductivity $\sigma_{xy}$ is a sensitive probe of FSR in YBCO, so is the thermal Hall conductivity $\kappa_{xy}$. Indeed, in the $T$ = 0 limit, the two are related by the Wiedemann-Franz law, whereby $\kappa_{xy} / T = L_0\, \sigma_{xy}$, with $L_0 = (\pi^2/3)(k_B/e)^2$, as recently verified in YBCO at $p$ = 0.11 for $H > H_{c2}$ (ref. 19). In other words, the large negative $\kappa_{xy}$ measured at low $T$ and high $H$ is perfectly consistent with the large negative $\sigma_{xy}$ (or $R_H = \rho_{xy} / H$) that first revealed broken translational symmetry in the field-induced normal state of underdoped YBCO (ref. 3). The advantage of $\kappa_{xy}$ over $\sigma_{xy}$ (or any other DC electric or thermo-electric coefficient) is that it can be used inside the superconducting state.

In a *d*-wave superconductor, $\kappa_{xy}$ is due to thermally excited quasiparticles outside the vortex cores, and it reflects the properties of the underlying Fermi surface[20,21] – in particular its curvature, which dictates the sign of the (thermal and electrical) Hall response.

In Fig. 2a, we show the temperature dependence of $\kappa_{xy}$ in an overdoped YBCO sample with $p = 0.18$ at $H = 3$ T, plotted as $\kappa_{xy}$ vs $T / T_c$, where $T_c$ is the superconducting critical temperature at $H = 3$ T (Fig. S1). At $p = 0.18$, there are no CDW modulations[16,17], and there is no indication of any broken symmetry or pseudogap, *e.g*. the upper critical field $H_{c2}$ has not yet started to drop[18], so the Fermi surface is expected to be the simple large hole-like cylinder characteristic of overdoped cuprates[22], with a positive normal-state Hall effect ($\sigma_{xy} > 0$). The vanishingly small value of $\kappa_{xy}$ above $T_c$ is due to the strong inelastic scattering in the normal state[23]. As soon as $T$ falls below $T_c$, $\kappa_{xy}$ undergoes a huge enhancement, because the opening of the superconducting gap wipes out much of the inelastic electron-electron scattering, causing the quasiparticle mean free path to grow by two orders of magnitude in clean samples[23]. When the mean free path reaches its upper limit imposed by elastic (defect) scattering, further cooling causes $\kappa_{xy}$ to fall, since the density of thermally excited quasiparticles decreases as $T \rightarrow 0$ (ref. 21). The two compensating effects lead to a peak, also seen in the longitudinal thermal conductivity $\kappa_{xx}(T)$ (ref. 24) and in the microwave conductivity $\sigma_1(T)$ (ref. 25). Note that $\kappa_{xy} > 0$ at all $T$, reflecting the positive curvature of the underlying Fermi surface.

We now turn to $\kappa_{xy}$ in the underdoped regime – never explored before in any cuprate, to the best of our knowledge. In Fig. 2a, we compare $\kappa_{xy}(T)$ at $p = 0.11$ and 0.12 with our reference data at $p = 0.18$, all normalized by their peak value. The three curves are almost identical: they have the same shape and they are all positive (except very near $T_c$ – see below). This shows that the underlying Fermi surface at $p = 0.11$ and 0.12 is not reconstructed at low field. It may be gapped in the anti-nodal regions by the pseudogap, but it must have essentially the same curvature as the original large cylinder. We conclude that long-range translational symmetry is not broken in underdoped YBCO at $T \to 0$ and $H = 0$, for $p = 0.11$-0.12. Given that all properties related to the FSR (refs. 4,7) and to the CDW modulations[16,17] evolve smoothly from $p \sim 0.08$ to $p \sim 0.16$, we infer that our conclusion holds throughout that doping range.

The shallow dip in $\kappa_{xy}(T)$ to negative values near $T_c$ (Fig. 2a) is investigated in Fig. S1, where we plot $\kappa_{xy}$ vs $T$ at $p = 0.12$ for various values of $H$. Upon cooling below 100 K, $\kappa_{xy}(T)$ decreases smoothly and changes sign at $T \sim 68$ K, for any field. The same sign change occurs in $R_H(T)$ (ref. 3), also at $T \sim 68$ K. The negative sign is a signature of the FSR caused by the short-range CDW modulations, in a regime where the electronic mean free path is known to be very short. Upon crossing below $T_c$, $\kappa_{xy}$ immediately reverts to being positive, showing that this low-field FSR is only observed when the mean free path is short. When it becomes long, it averages over the short-range CDW modulations, and translational symmetry is not broken. We therefore arrive at our first main finding: translational symmetry is not broken in the zero-field ground state of underdoped YBCO, at least down to $p \sim 0.08$ (Fig. 1).

In Fig. 2b, we see that the positive $\kappa_{xy}(T)$ at low $H$ becomes negative at high $H$. This implies that FSR happens at some intermediate field, before reaching the upper critical field $H_{c2} = 24 \pm 1$ T (refs. 18,19). We can pinpoint the onset field for FSR by looking at field sweeps. In Fig. 3a, we show the field dependence of $\kappa_{xy}$ in YBCO at $p = 0.11$ and 0.18, plotted as $\kappa_{xy} / T$ vs $H$, for $T < T_c$. The standard behavior is exhibited by our reference overdoped sample with $p = 0.18$ : a rapid initial rise in $\kappa_{xy}$ vs $H$ at very low $H$ is soon followed by a peak and then a slow decrease, saturating to a small positive value at high $H$, in agreement with previous work[23] (Fig. S2). The same behavior is observed in the underdoped samples, but only up to a certain field. At $p = 0.11$, the behavior characteristic of the overdoped regime stops at $H = 18$ T, as seen in Fig. 3a for an isotherm at $T = 22$ K. Above 18 T, $\kappa_{xy}$ drops to become negative. This drop is the clear signature of FSR, occurring at an onset field we label $H_{FSR}$. Isotherms at various temperatures (Figs. 3 and 4, and Fig. S3) allow us to determine $H_{FSR}$ as a function of temperature, plotted on an $H$-$T$ diagram in Fig. 5a. We observe that $H_{FSR}$ is constant from $T \sim 0$ up to $T \sim 20$ K, with $H_{FSR} = 18 \pm 1$ T at $p = 0.11$. Similar measurements on our sample with $p = 0.12$ (Fig. 3 and Fig. S3) also yield a sharp onset field for FSR, again flat at low temperature, but with a slightly lower value: $H_{FSR} = 15 \pm 1$ T at $p = 0.12$ (Fig. 5b). (Note that $H_{FSR}$ cannot be tracked above ~ 20 K since these higher temperatures are getting close to $T_c(H)$ and the associated regime of strong inelastic scattering and short mean free path; see Fig. S4.) The existence of a threshold field for FSR is our second main finding. It resolves the apparent contradiction between the Fermi surface detected by quantum oscillations – a closed electron-like pocket – and that observed by ARPES (ref. 26) – an open arc – by showing that the former is measured at $H > H_{FSR}$ while the latter is measured at $H < H_{FSR}$ .

In the $H$-$T$ diagram of Fig. 5a, we see that the onset of FSR at $H_{FSR} = 18$ T for $p = 0.11$ coincides exactly with $H_{ultrasound} = 18$ T, the field at which a phase transition is detected in the sound velocities at $p = 0.11$ ($T_c = 61$ K), constant below ~ 35 K (ref. 10). We therefore arrive at our third main finding: the field-induced thermodynamic transition seen in ultrasound is where long-range translational symmetry is broken. Recent XRD measurements in pulsed fields up to 28 T have revealed that long-range 3D CDW order appears above $H$ = 15 T in YBCO at $p$ = 0.12 (ref. 27), a threshold field in excellent agreement with our $H_{FSR}$ at $p$ = 0.12 (Fig. 5b).

Although the transition seen in ultrasound has been attributed to the onset of charge order detected by NMR, the two sharp onset fields are clearly different (Fig. 5a). At the same doping, $p$ = 0.11 ($T_c$ = 61 K), $H_{ultrasound}$ = 18 ± 0.5 T (ref. 10), while $H_{NMR}$ = 10.4 ± 1.0 T (ref. 9). Our explanation for this discrepancy is that NMR detects CDW inside the core of vortices before a bulk transition has taken place. Our evidence to support this scenario is shown in Fig. 4, where we compare the thermal and electrical Hall signals measured in the same YBCO sample (with $p$ = 0.11) at the same temperature ($T$ ~ 15 K). As a function of increasing $H$, the electrical Hall coefficient $R_H(H)$ is zero up to the vortex-solid melting field $H_{vs}(T)$ (blue line in Fig. 5a), beyond which it becomes non-zero due to flux flow (the movement of vortices). By contrast, thermal transport ($\kappa_{xx}$ and $\kappa_{xy}$) is insensitive to flux flow. We see that $R_H$ is negative above $H$ ~ 10 T, well below the field at which $\kappa_{xy}$ becomes negative (Fig. 4), namely $H_{FSR}$ = 18 T (Fig. 5a). This difference must be due to the vortices, implying that FSR first occurs inside the vortex core. Electrical transport reflects the state in the core and, as a local probe, NMR can also detect it. This is reminiscent of the modulations detected around vortices in cuprate superconductors via STM (ref. 28). Note that for the two dopings available, $H_{FSR}$ roughly scales with $H_{NMR}$ (Fig. S5).

Having determined the onset field for FSR, we now ask what is the onset temperature for FSR. For this, we use electrical Hall data from our YBCO sample with $p$ = 0.11. In Fig. S6, we plot $R_H(T)$ at $H$ = 10 T < $H_{FSR}$ and at $H$ = 35 T > $H_{FSR}$. Upon cooling, $R_H(H$ = 10 T) decreases slowly and monotonically from 80 K down to 40 K. We attribute this slow drop to the effect of short-range CDW modulations, whose correlation length is greater than (or comparable to) the electronic mean free path above $T_c(H)$. By contrast, $R_H(H = 35$ T) drops abruptly below 60 K, in excellent agreement with the sharp onset of charge order at $T_{NMR} = 57 \pm 5$ K, seen by NMR (in a field of 28 T) for $p = 0.11$ (ref. 9) (Fig. 5a). We conclude that NMR, ultrasound and thermal Hall transport all detect the same field-induced long-range charge order.

The *H-T* phase diagram of FSR in YBCO is shown in Fig. 5. It reveals a scenario of strong phase competition between long-range CDW order, bounded by the red line, and long-range superconductivity, bounded by the green line. Superconductivity completely excludes CDW order at low field, up to $H_{FSR} = H_{ultrasound}$. Note that $H_{FSR}$ lies distinctly below the upper critical field $H_{c2}$ where the vortex phase ends, as detected by high-field measurements of $\kappa_{xx}$ (ref. 18) and $\kappa_{xy}$ (ref. 19). So there is a region of phase coexistence. Both $H_{FSR}(T)$ and $H_{c2}(T)$ are flat below $T \sim 20$ K, with $H_{FSR} / H_{c2} = 0.75$ at $p = 0.11$ (Fig. 5a) and 0.63 at $p = 0.12$ (Fig. 5b). The separation between $H_{FSR}$ and $H_{c2}$ is a measure of the strength of phase competition. The fact that it is wider at $p = 0.12$ than at $p = 0.11$ (for the same $H_{c2} \sim 24$ T) implies that CDW order is stronger at $p = 0.12$, consistent with the fact that $T_{NMR}$ is highest at $p = 0.12$ (Fig. 1).

In the absence of superconductivity (*i.e.* at high fields), there is a critical doping $p_{CDW}$ below which long-range CDW order sets in at $T = 0$. Previous Hall data show that $p_{CDW} > 0.15$ (ref. 4). Our findings demonstrate that this $T = 0$ critical point does not exist in zero field – the competition from superconductivity is so intense that it wipes it out. This could explain why the enhancement of the quasiparticle mass measured recently in the field-induced normal state of YBCO (ref. 29) as $p \to 0.18$ is not manifest in zero-field properties, in a scenario where the mass enhancement is associated with the CDW quantum critical point.

Although the correlation length of the XRD-detected CDW modulations at $H = 0$ and $T \rightarrow 0$ is too short to break translational symmetry, these modulations could nevertheless break rotational symmetry, below $T_{XRD}$, given that they are known to be anisotropic[16,17,30,31]. In that case, $p_{nem} \sim 0.16$ (ref. 17) would be the critical doping for the onset of nematicity in YBCO (Fig. 1), and possibly a nematic quantum critical point[32]. This would be qualitatively consistent with STM studies on $Bi_2Sr_2CaCu_2O_{8+\delta}$, which find rotational and translational symmetries to be broken on long and short length scales, respectively, below a common critical doping $p_{nem} \sim 0.19$ (ref. 33). The onset of nematicity in YBCO below $p_{nem} \sim 0.16$ at $T = 0$ is also consistent with the spontaneous onset of in-plane anisotropy in the microwave conductivity $\sigma_1(T)$ of YBCO at $T << T_c$ , occurring between $p = 0.185$, where $\sigma_{1b} / \sigma_{1a} \sim 1$, and $p = 0.10$, where $\sigma_{1b} / \sigma_{1a} \sim 3$ (ref. 25).

From a general perspective, our study demonstrates how the thermal Hall effect can be a powerful probe of phase transitions inside the superconducting state, such as those that occur in organic, heavy-fermion, and iron-based superconductors.

## Acknowledgements

L.T. thanks ESPCI-ParisTech, Université Paris-Sud, CEA-Saclay and the Collège de France for their hospitality and support, and the École Polytechnique (ERC-319286 QMAC) and LABEX PALM (ANR-10-LABX-0039-PALM) for their support, while this article was written. A portion of this work was performed at the Laboratoire National des Champs Magnétiques Intenses, which is supported by the French ANR SUPERFIELD, the EMFL, and the LABEX NEXT. Another portion of this work was performed at the National High Magnetic Field Laboratory, which is supported by the National Science Foundation Cooperative Agreement No. DMR-1157490, the State of Florida, and the U.S. Department of Energy. R.L., D.A.B. and W.N.H. acknowledge funding from the Natural Sciences and Engineering Research Council of Canada (NSERC). L.T. acknowledges support from the Canadian Institute for Advanced Research (CIFAR) and funding from NSERC, the Fonds de recherche du Québec - Nature et Technologies (FRQNT), the Canada Foundation for Innovation (CFI) and a Canada Research Chair.

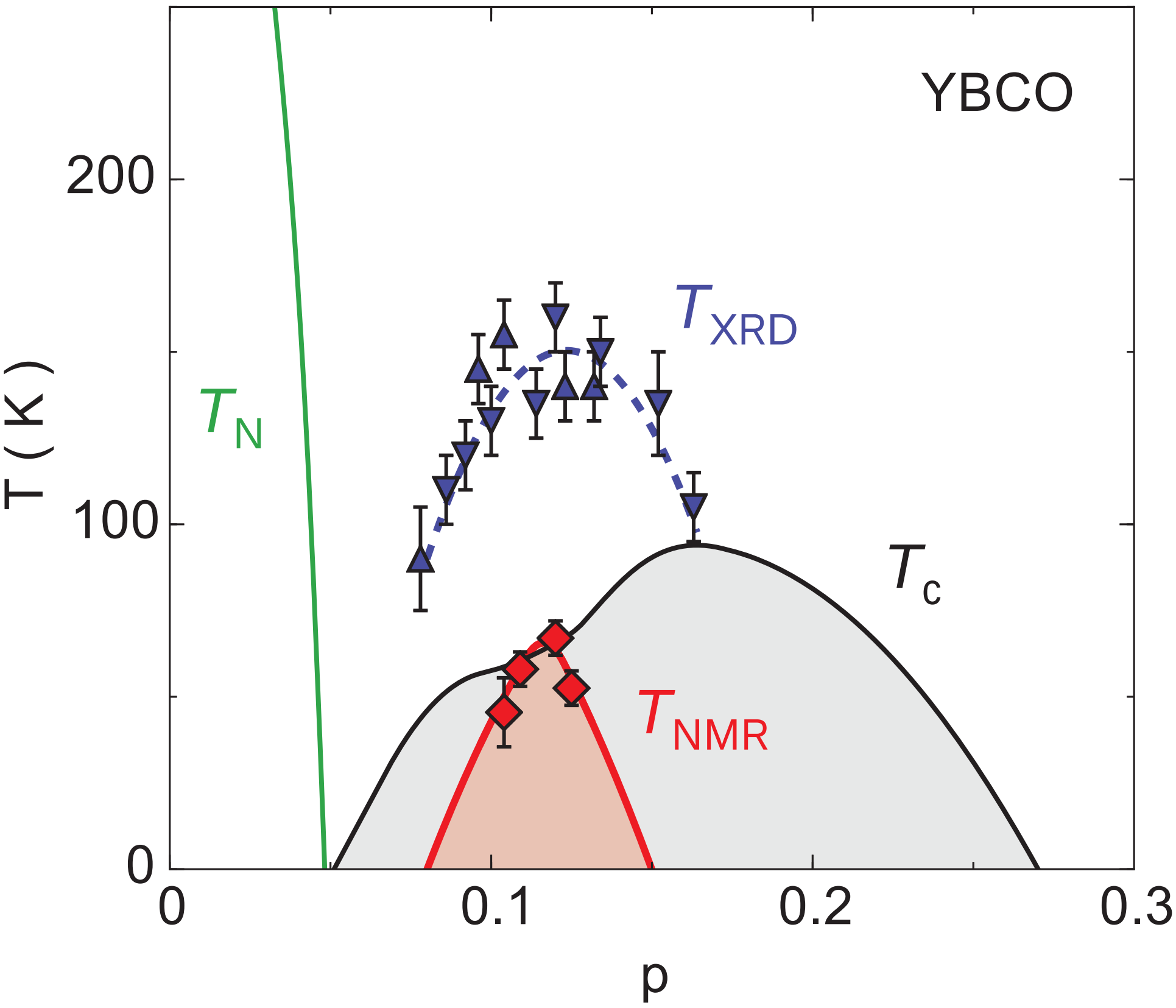

**Figure 1 | Temperature-doping phase diagram.**

Temperature-doping phase diagram of YBCO, with the superconducting phase (grey dome) below $T_c$ (black line) and the antiferromagnetic phase below $T_N$ (green line). Long-range charge order (red dome) is detected by NMR (ref. 8) below a transition temperature $T_{NMR}$ (red diamonds from ref. 9), but only above a threshold magnetic field. Also shown is the onset of short-range charge-density-wave modulations detected by X-ray diffraction below $T_{XRD}$ (up triangles from ref. 16; down triangles from ref. 17). Red and blue lines are a guide to the eye.

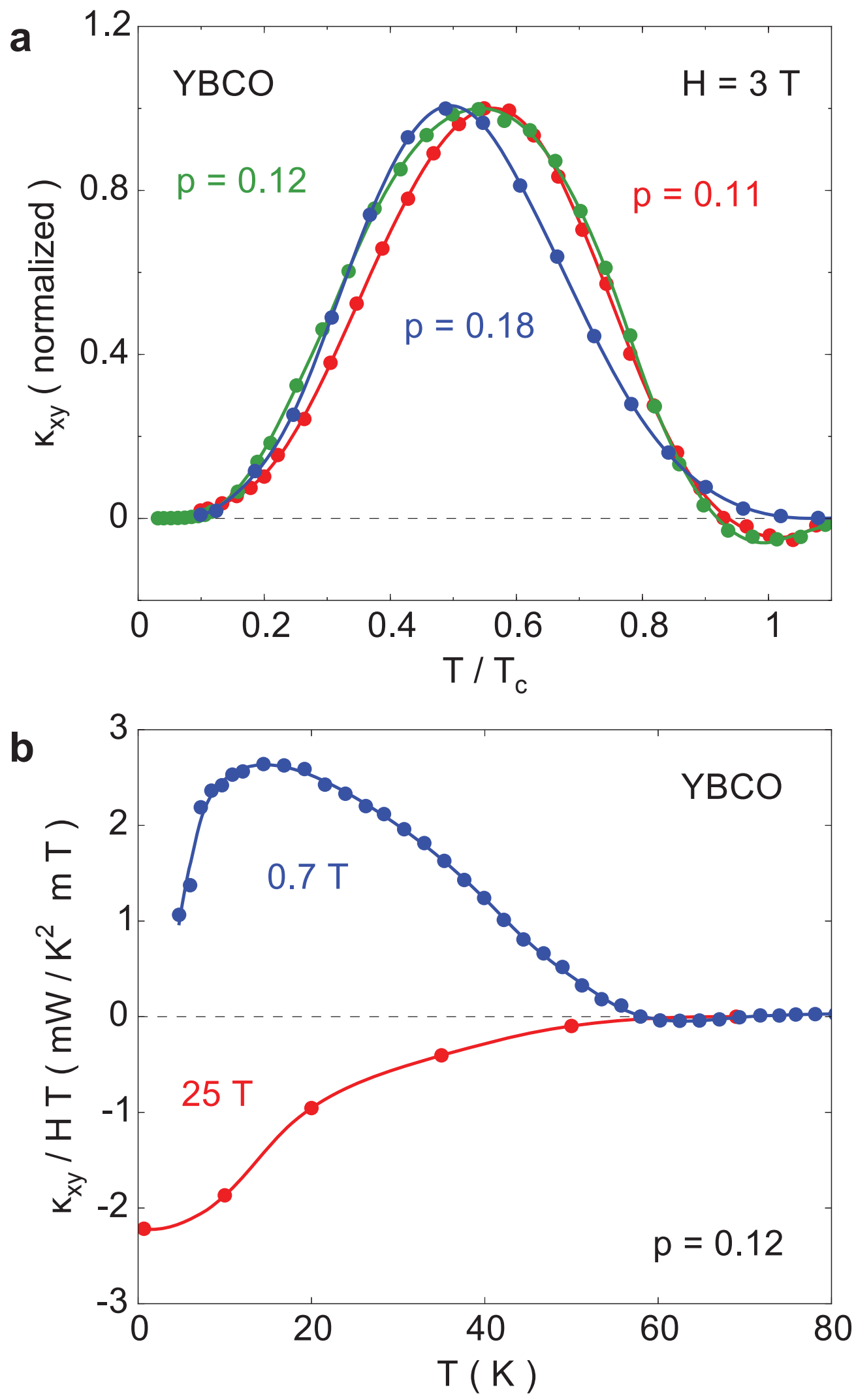

**Figure 2 | Thermal Hall conductivity *vs* temperature.**

**a)** Thermal Hall conductivity $\kappa_{xy}$ measured in a magnetic field $H$ = 3 T as a function of temperature in three single crystals of YBCO with dopings $p$ as indicated, plotted as $\kappa_{xy}$ vs $T$ / $T_c$ , where $T_c$ is the critical temperature at $H$ = 3 T (Fig. S1). $\kappa_{xy}$ is normalized so that its peak value is 1.0. The horizontal dashed line marks $\kappa_{xy}$ = 0. **b)** Thermal Hall conductivity of YBCO at $p$ = 0.12 for two values of the magnetic field, $H$ = 0.7 T (blue) and $H$ = 25 T (red), plotted as $\kappa_{xy}$ / ($H$ $T$) vs $T$. The data at $H$ = 25 T correspond to the normal state, since $H = H_{c2}$ (ref. 18), and it satisfies the Wiedemann-Franz law at $T \rightarrow 0$ (ref. 19).

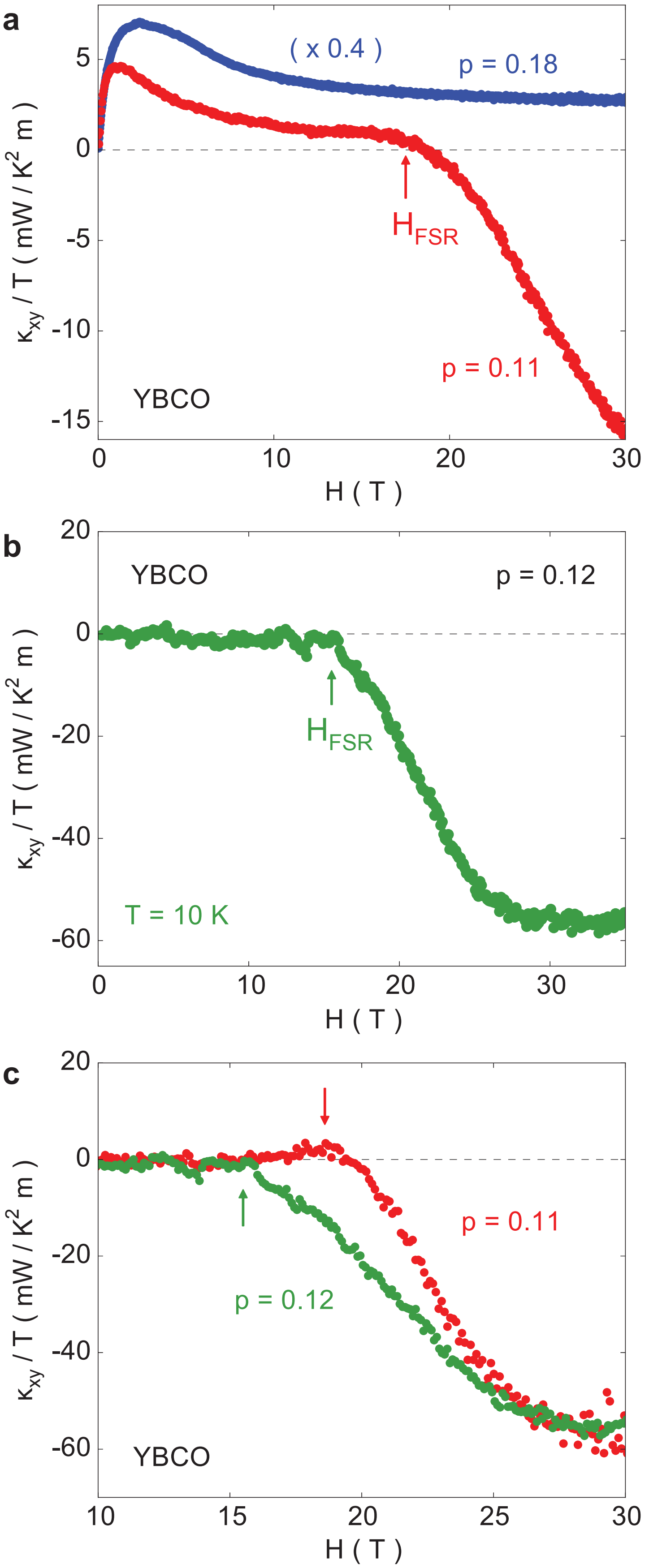
a
( x 0.4 )
p = 0.18
$H_{FSR}$
p = 0.11
YBCO
$\kappa_{xy}$ / T ( mW / K$^2$ m )
H ( T )
b
YBCO
p = 0.12
$H_{FSR}$
T = 10 K
$\kappa_{xy}$ / T ( mW / K$^2$ m )
H ( T )
c
p = 0.11
p = 0.12
YBCO
$\kappa_{xy}$ / T ( mW / K$^2$ m )
H ( T )

**Figure 3 | Thermal Hall conductivity *vs* magnetic field.**

**a)** Thermal Hall conductivity $\kappa_{xy}$ as a function of magnetic field *H* in two single crystals of YBCO with dopings *p* as indicated, plotted as $\kappa_{xy}$ / *T* vs *H.* The data at *p* = 0.11 (red) and *p* = 0.18 (blue; multiplied by a factor 0.4) were taken at *T* = 22 K and *T* = 35 K, respectively. **b)** Same for *p* = 0.12, at *T* = 10 K. **c)** Comparison of isotherms at two dopings: *p* = 0.11 (red; *T* = 8 K) and *p* = 0.12 (green; *T* = 10 K). In all panels, the arrows mark the field $H_{FSR}$ at which $\kappa_{xy}$ starts to fall towards negative values. This is the onset field for Fermi-surface reconstruction. Panel C shows that $H_{FSR}$ is larger at *p* = 0.11 than at *p* = 0.12.

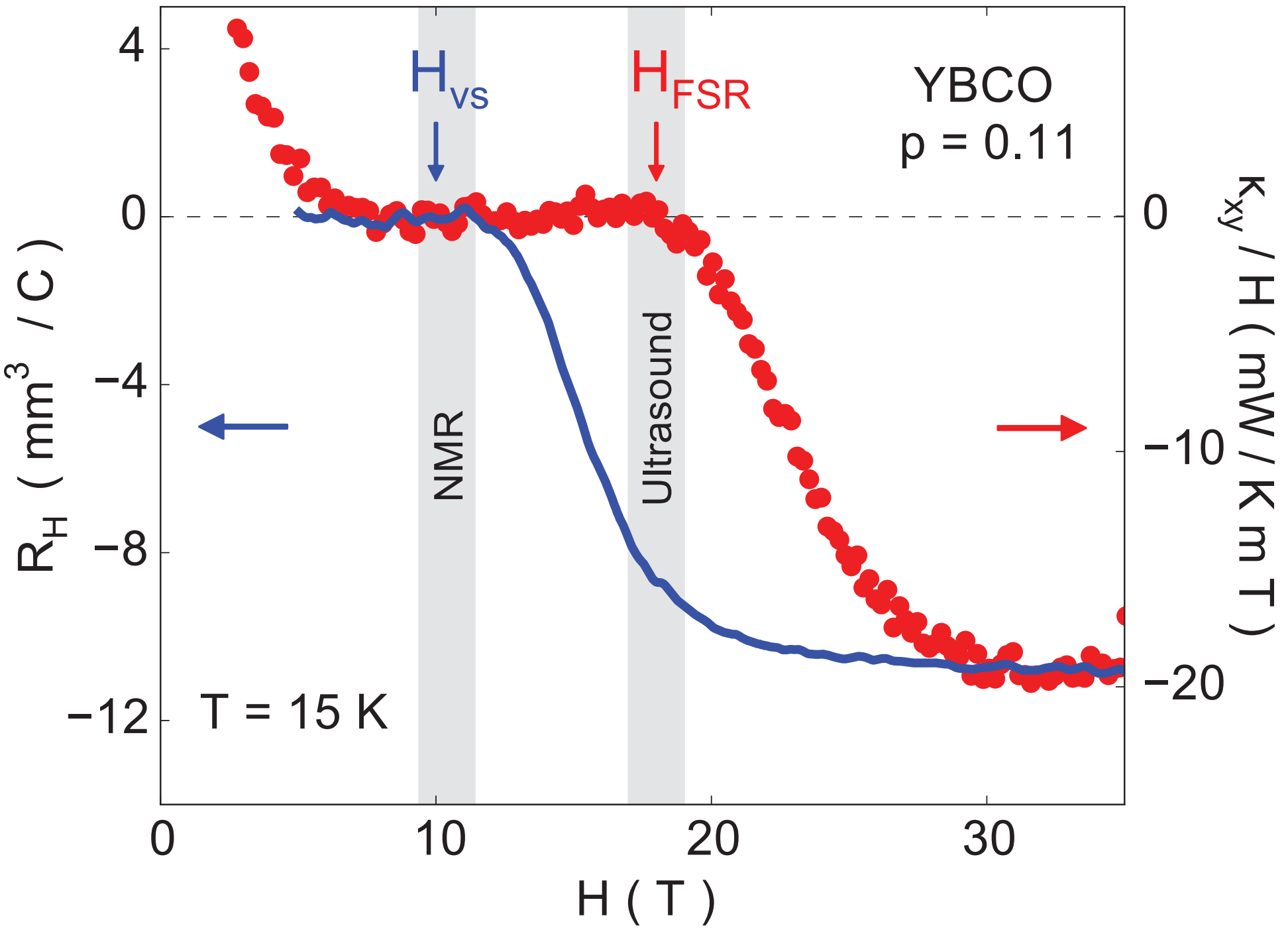

**Figure 4 | Comparing thermal and electrical Hall conductivities.**

Thermal Hall conductivity $\kappa_{xy}$, plotted as $\kappa_{xy}$ / $H$ (red; right axis), and electrical Hall coefficient $R_H$ (blue; left axis), as a function of field $H$, measured in the same single crystal of YBCO with $p$ = 0.11, at $T$ = 16 K and 14 K, respectively. The onset of Fermi-surface reconstruction detected in $\kappa_{xy}$ at $H = H_{FSR}$ = 18 T (red arrow) coincides with the thermodynamic transition detected in ultrasound measurements (upper grey band)[10]. By contrast, $R_H$ drops to become negative above $H$ ~ 10 T, a field which coincides with the onset of CDW order detected in NMR measurements (lower grey band)[9]. (At higher temperature, $R_H(H)$ is initially positive above $H_{vs}$ , and it starts to drop towards negative values at $H$ ~ 10 T.) We attribute this lower onset field to CDW order inside the vortex cores (see text). The movement of these cores in the flux-flow regime above the vortex-solid melting field $H_{vs}$ (blue arrow; Fig. 5a) is detected in electrical transport, but not in thermal transport.

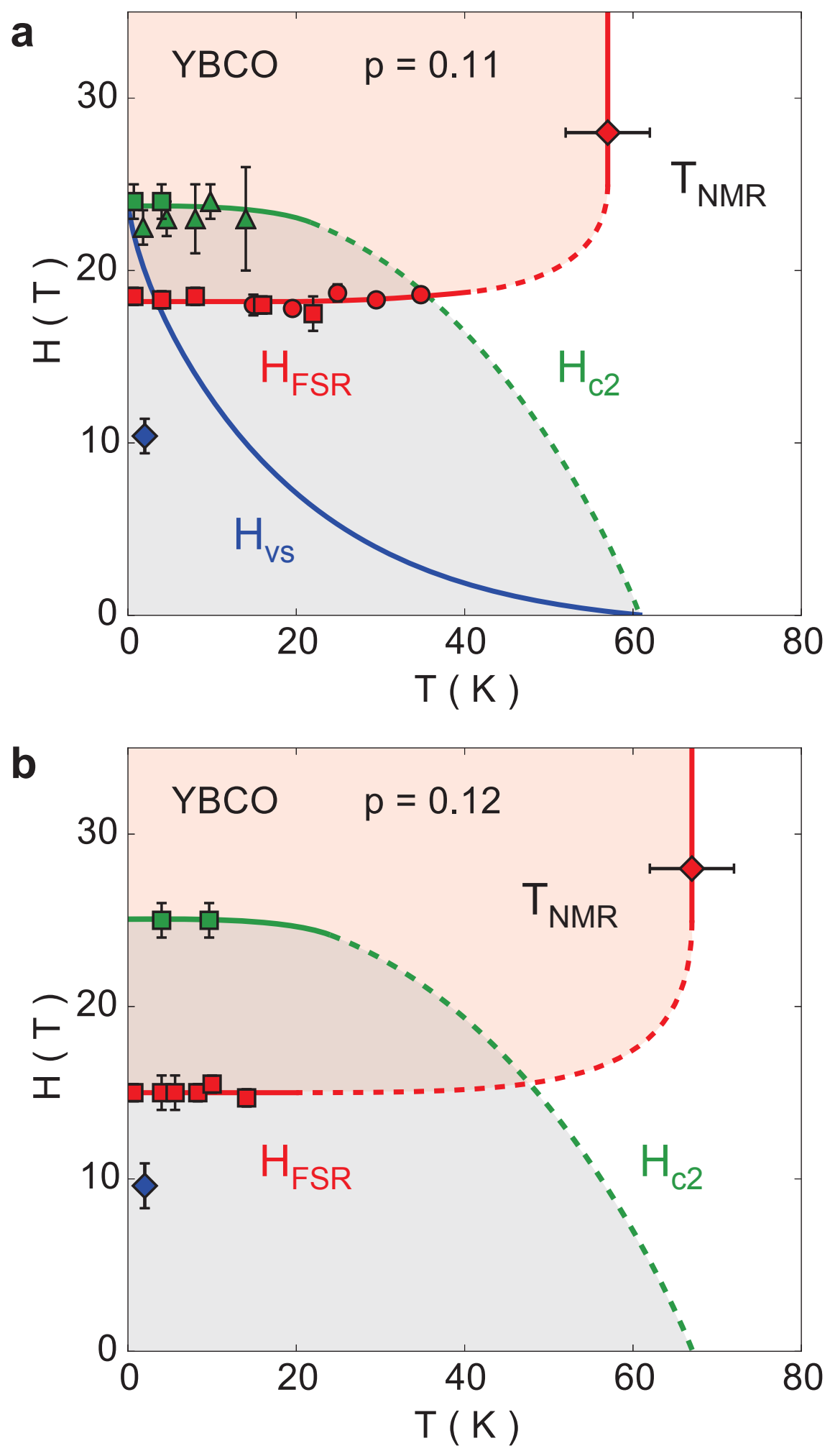

**Figure 5 | Magnetic field-temperature phase diagram.**

**a)** Magnetic field-temperature phase diagram of YBCO for $p$ = 0.11, showing the vortex-solid melting field $H_{vs}$ (blue line; from ref. 18), the upper critical field $H_{c2}$ (green; triangles from ref. 18, squares from ref. 19), and the onset field for Fermi-surface reconstruction, $H_{FSR}$ (red squares; from Figs. 3, 4 and S3). Also shown is the critical field for the thermodynamic transition detected in the sound velocity of YBCO with $p$ = 0.11 (red circles, from ref. 10), and the onset field $H_{NMR}$ (blue diamond) and temperature $T_{NMR}$ (red diamond) for CDW order detected as a splitting of the NMR line in YBCO with $p$ = 0.11 (ref. 9). **b)** Same for $p$ = 0.12, leaving out the $H_{vs}$ line.

# SUPPLEMENTARY MATERIAL

## Onset field for Fermi-surface reconstruction in the cuprate superconductor $YBa_2Cu_3O_y$

### SAMPLES

Single crystals of $YBa_2Cu_3O_y$ (YBCO) were obtained by flux growth, as described in ref. 34. Our samples are detwinned single crystals of $YBa_2Cu_3O_y$ with oxygen content $y$ = 6.54, 6.67, and 6.998, with a high degree of oxygen order (ortho-II, ortho-VIII, and ortho-I, respectively). The hole doping $p$ is obtained from the superconducting $T_c$ (ref. 34), defined as the temperature where the electrical resistance goes to zero. Our samples had a $T_c$ of 60.5 K, 67 K, and 90.5 K, giving $p$ = 0.11, 0.12, and 0.18, respectively. The samples are rectangular platelets with six contacts applied in the standard geometry, using diffused gold pads.

### MEASUREMENT OF THE THERMAL HALL CONDUCTIVITY

The thermal Hall conductivity $\kappa_{xy}$ of our YBCO samples was measured in Sherbrooke up to 18 T, at the LNCMI in Grenoble up to 34 T, and at the NHMFL in Tallahassee up to 35 T. In all measurements, the magnetic field was applied along the $c$ axis, normal to the $CuO_2$ planes.

The measurement procedure was described in detail in ref. 19, where we reported a thermal Hall study of YBCO in the normal state, at $H > H_{c2}$. Here, our emphasis is on the superconducting state, for $H < H_{c2}$. For the present study, in addition to field sweeps of $\kappa_{xy}$ vs $H$ at fixed temperature $T$, we also performed measurements of $\kappa_{xy}$ vs $T$, at fixed magnetic field (*e.g.* Figs. 2, S1, S4). The data obtained in the two different ways ($H$ sweeps and $T$ sweeps) agree very well.

### PRIOR MEASUREMENTS OF THE THERMAL HALL CONDUCTIVITY IN CUPRATES

In a recent study, we used the thermal Hall conductivity to test the Wiedemann-Franz law in the normal state of underdoped YBCO, at magnetic fields greater than $H_{c2}$ (ref. 19). We did not investigate the superconducting state, the focus of the present work.

Apart from that study, all previous thermal Hall effect measurements in cuprates were performed on optimally doped or overdoped samples, with $p$ = 0.16 - 0.18 (refs. 23, 35, 36, 37, 38). Their main common finding was that $\kappa_{xy}$ vs $T$ peaks below $T_c$, with a peak position approximately at $T_c$ / 2. This was analyzed convincingly in terms of a large enhancement of the electronic mean free path, due to a loss of inelastic scattering upon entering the superconducting state[23]. The peak is all the more pronounced in cleaner samples. It is much larger in YBCO than in $Bi_2Sr_2CaCu_2O_{8+\delta}$ (ref. 37), samples of the former being much more ordered than samples of the latter.

An estimate of the electronic mean free path in YBCO at $p$ = 0.18 gives ~ 10 nm at $T_c$ (ref. 23). Given that the correlation length of short-range CDW modulations in YBCO at $T_c$ is also approximately 10 nm (at $p$ = 0.11 - 0.12) (ref. 17), it is reasonable to expect that charge carriers will perceive the new periodicity of those CDW modulations just above $T_c$, and transport properties such as the Hall effect (thermal or electrical) will harbor signatures of FSR, *i.e.* $R_H < 0$ and $\kappa_{xy} < 0$. However, as soon as $T$ is lowered below $T_c$, the mean free path increases very rapidly, by a factor of ~ 100 (ref. 23), almost immediately exceeding the CDW correlation length. And indeed, we find that FSR signatures disappear, as $\kappa_{xy}$ immediately goes back to being positive (Fig. S1).

Thermal Hall measurements of optimally-doped YBCO were also used to obtain the Lorenz ratio of heat to charge conductivities in the normal state above $T_c$ (ref. 38), found to be $(\kappa_{xy} / T) / (L_0 \sigma_{xy})$ ~ 0.15 just above $T_c$. This showed that the strong inelastic scattering present in YBCO above $T_c$ suppresses heat conduction much more effectively than charge conduction.

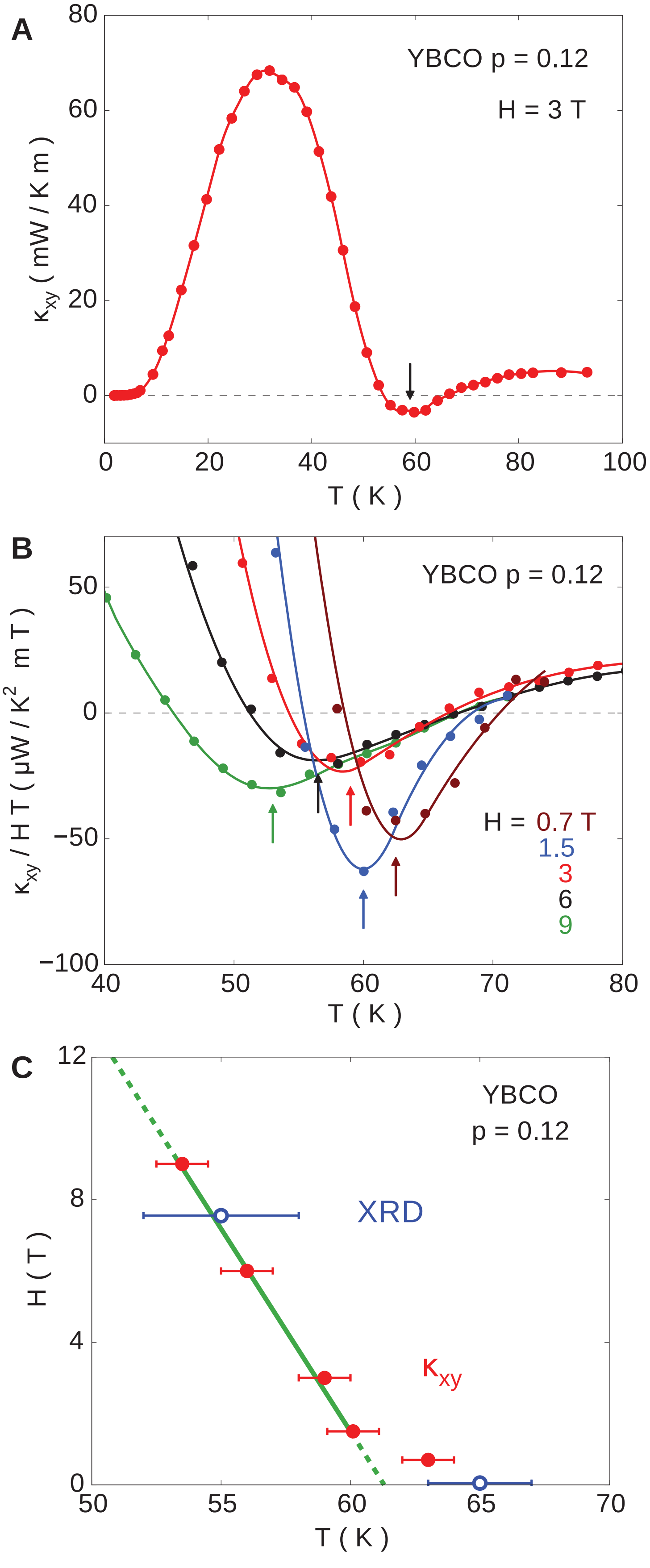

A
YBCO p = 0.12
H = 3 T
κxy ( mW / K m )
T ( K )
B
YBCO p = 0.12
H = 0.7 T
1.5
3
6
9
κxy / H T ( μW / K² mT )
T ( K )
C
YBCO
p = 0.12
XRD
Kxy
H ( T )
T ( K )

**Figure S1 | Detection of $T_c$ in thermal Hall measurements.**

**a)** $\kappa_{xy}$ vs *T* in YBCO at *p* = 0.12, and *H* = 3 T. Upon cooling from 100 K, the slow drop in $\kappa_{xy}$ to negative values is a signature of the FSR caused by the short-range CDW modulations which develop below $T_{XRD}$ (Fig. 1), in a regime where the electronic mean free path is very short. When the mean free path suddenly increases below $T_c$ (ref. 23), $\kappa_{xy}$ immediately reverts to being positive. This shows that the Fermi surface is not truly reconstructed at those low fields. A long mean free path below $T_c$ averages over the short CDW correlation length, and translational symmetry is not broken on a long length scale. **b)** Temperature dependence of $\kappa_{xy}$ in YBCO at *p* = 0.12, plotted as $\kappa_{xy}$ / (*H T*) vs *T* for different fields as indicated. Arrows indicate the location of the minimum. **c)** Location of the minima in $\kappa_{xy}$ / *T* vs *T* in the *H*-*T* plane, from data in panel B (red circles). The superconducting transition temperature $T_c$ is given for two field values (open blue circles), *H* = 0 and *H* = 7.5 T, obtained from X-ray measurements of the CDW intensity vs *T* in YBCO at *p* = 0.12 (ref. 12) (XRD). The X-ray intensity grows smoothly with decreasing temperature[16,17] until it is suddenly curtailed by the onset of superconductivity at $T = T_c$, due to strong phase competition. This produces a sharp cusp in the intensity vs *T*, located precisely at $T_c$ . The temperature at which the cusp is observed is therefore an accurate measure of $T_c(H)$. The consistency found here between red and blue data points shows that the minimum in $\kappa_{xy}$ is indeed located at $T_c(H)$.

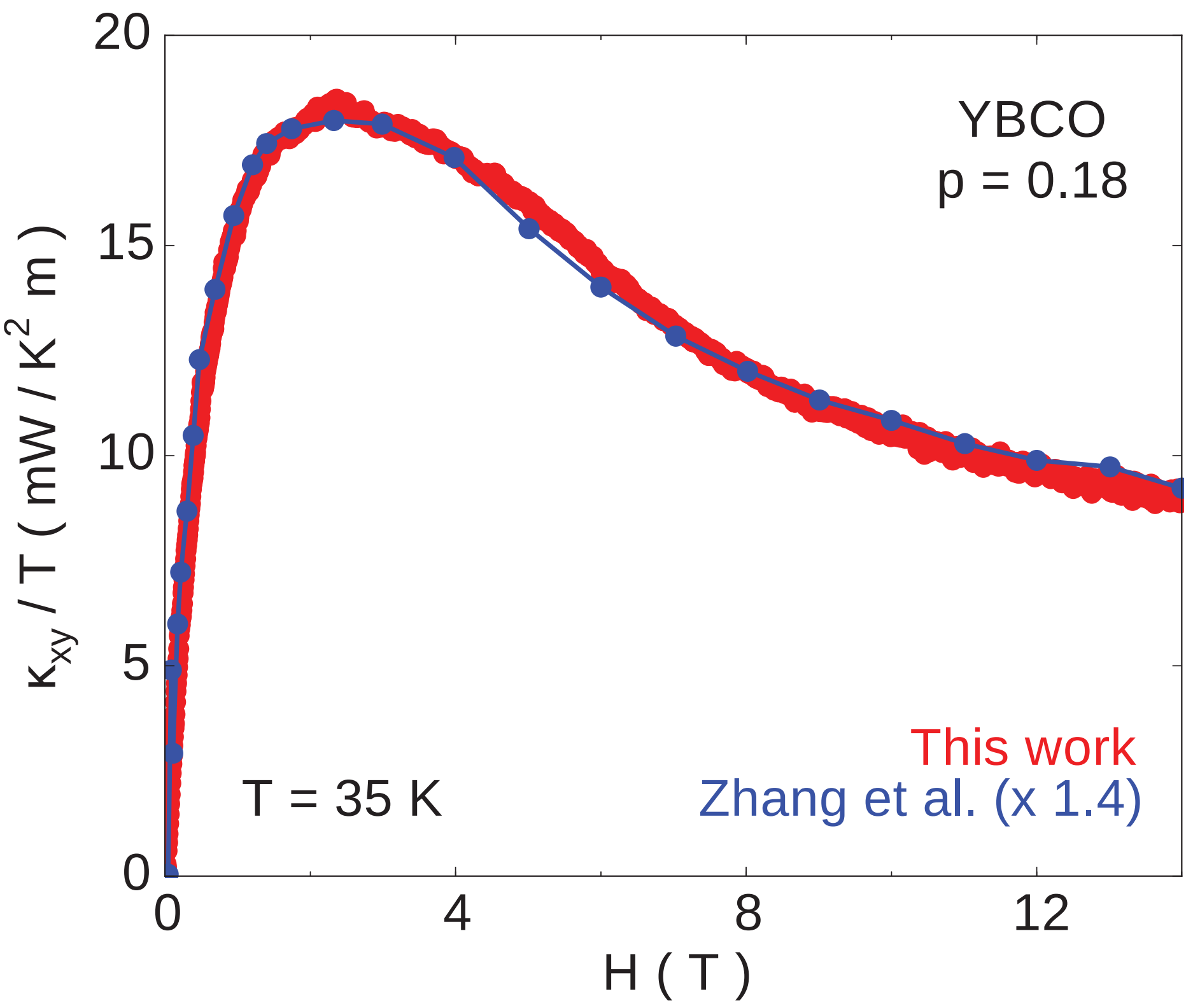

**Figure S2 | Comparison to prior measurements on overdoped YBCO samples.**

Thermal Hall conductivity of overdoped YBCO at $p$ = 0.18, plotted as $\kappa_{xy}$ vs $H$, at $T$ = 35 K. Our data (red) is compared with previous measurements by Zhang *et al.* [23] (blue circles; multiplied by a factor 1.4). The two data sets are seen to be in excellent agreement. The small quantitative discrepancy (by a factor 1.4) probably comes from the longer elastic mean free path in our sample, due to a lower density of oxygen vacancies in our crystal, whose oxygen content is $y$ = 6.998, compared to the crystal used by Zhang *et al.*, where $y$ = 6.99 (ref. 23).

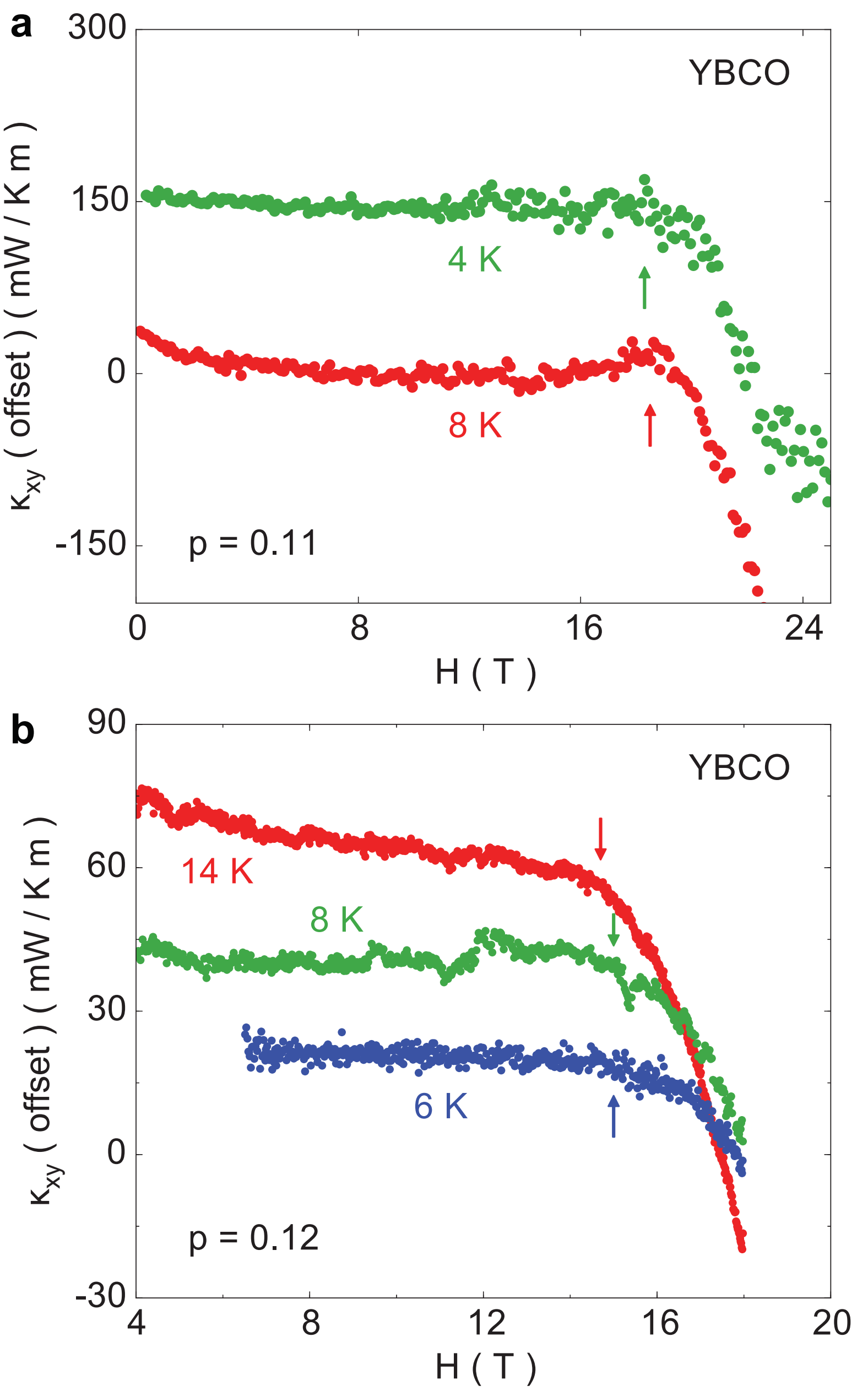

**Figure S3 | Isotherms of $\kappa_{xy}$ vs *H* in YBCO at *p* = 0.11 and *p* = 0.12.**

**a)** Thermal Hall conductivity of YBCO at *p* = 0.11, plotted as $\kappa_{xy}$ vs *H*, at temperatures as indicated. **b)** Same for *p* = 0.12. For the sake of clarity, the data are shifted rigidly upwards by different amounts, to separate the different isotherms. The arrows mark the location of $H_{FSR}$, the onset field for Fermi-surface reconstruction, plotted on the *H*-*T* phase diagrams of Fig. 5.

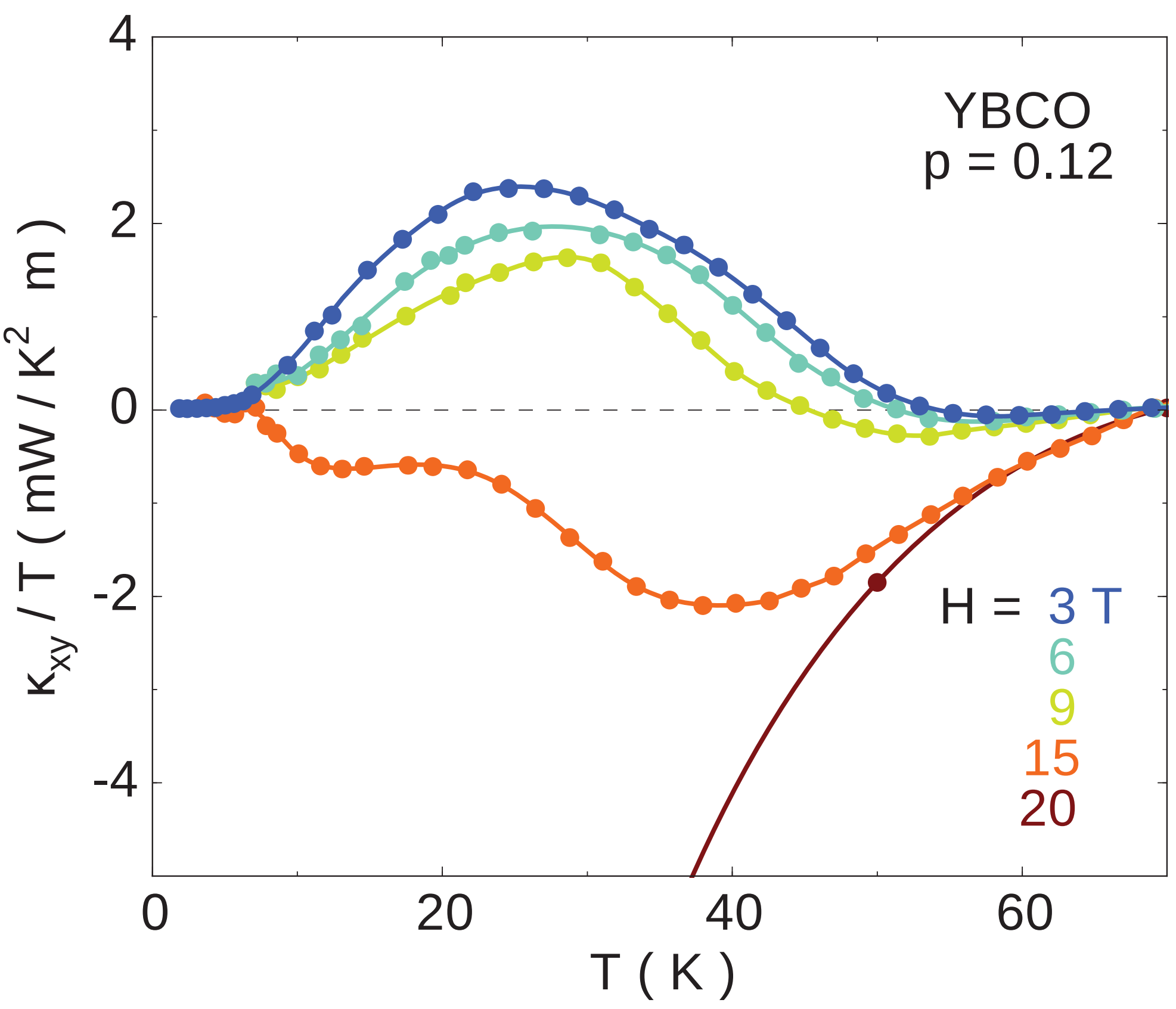

**Figure S4 | Thermal Hall conductivity of YBCO at *p* = 0.12.**

Thermal Hall conductivity of YBCO at $p$ = 0.12, plotted as $\kappa_{xy}$ / $T$ vs $T$, for different magnetic fields as indicated. As seen most clearly in the 15 T curve (orange), there are two regimes: 1) a low-$T$ regime below 20 K or so, where the electronic mean free path deep inside the superconducting state is very long; 2) a high-$T$ regime above 20 K or so, where the electronic mean free path is shorter, and very short above $T_c(H)$. In the low-$T$ regime, $\kappa_{xy}$ only becomes negative above a sharply defined onset field $H_{FSR}$ = 15 ± 1 T, which we attribute to the transition into a phase of long-range CDW order. In the high-$T$ regime, $\kappa_{xy}$ is negative just above $T_c(H)$ (minimum) at all fields (see Fig. S1), the result of an FSR caused by short-range CDW modulations, with a correlation length comparable or larger than the short mean free path.

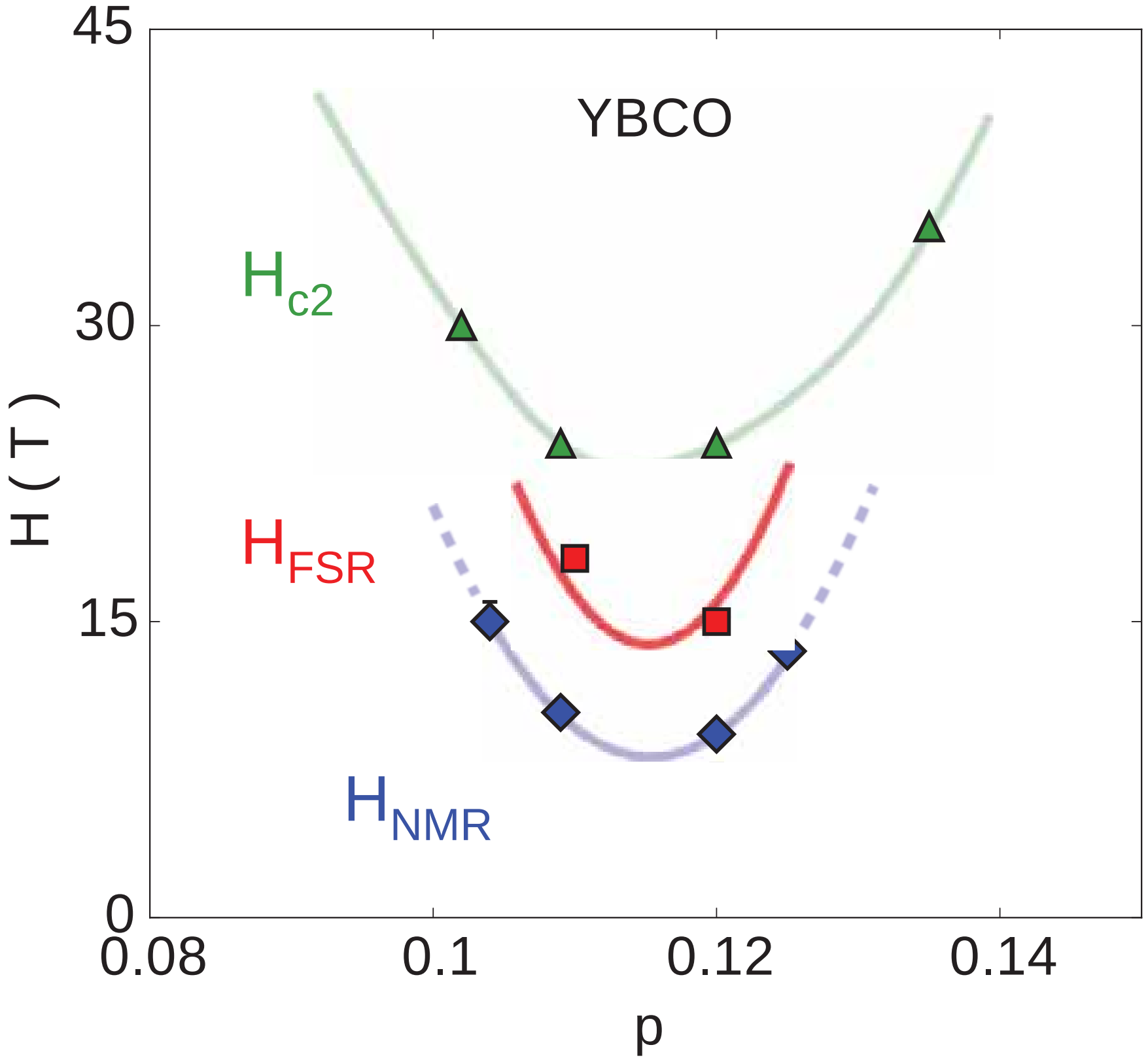

**Figure S5 | Doping dependence of the characteristic fields in YBCO.**

Comparison of three characteristic fields of YBCO in the *H*-*p* plane: 1) $H_{FSR}$ (red squares), the onset field for Fermi-surface reconstruction, detected in the thermal Hall conductivity $\kappa_{xy}$ vs *H* (Fig. 5); 2) $H_{NMR}$ (blue diamonds), the onset field for the splitting of the NMR line by charge order[9] ; 3) $H_{c2}$ (green triangles), the upper critical field for the end of the vortex state[18]. The three characteristic fields are obtained in the $T = 0$ limit. All lines are a guide to eye.

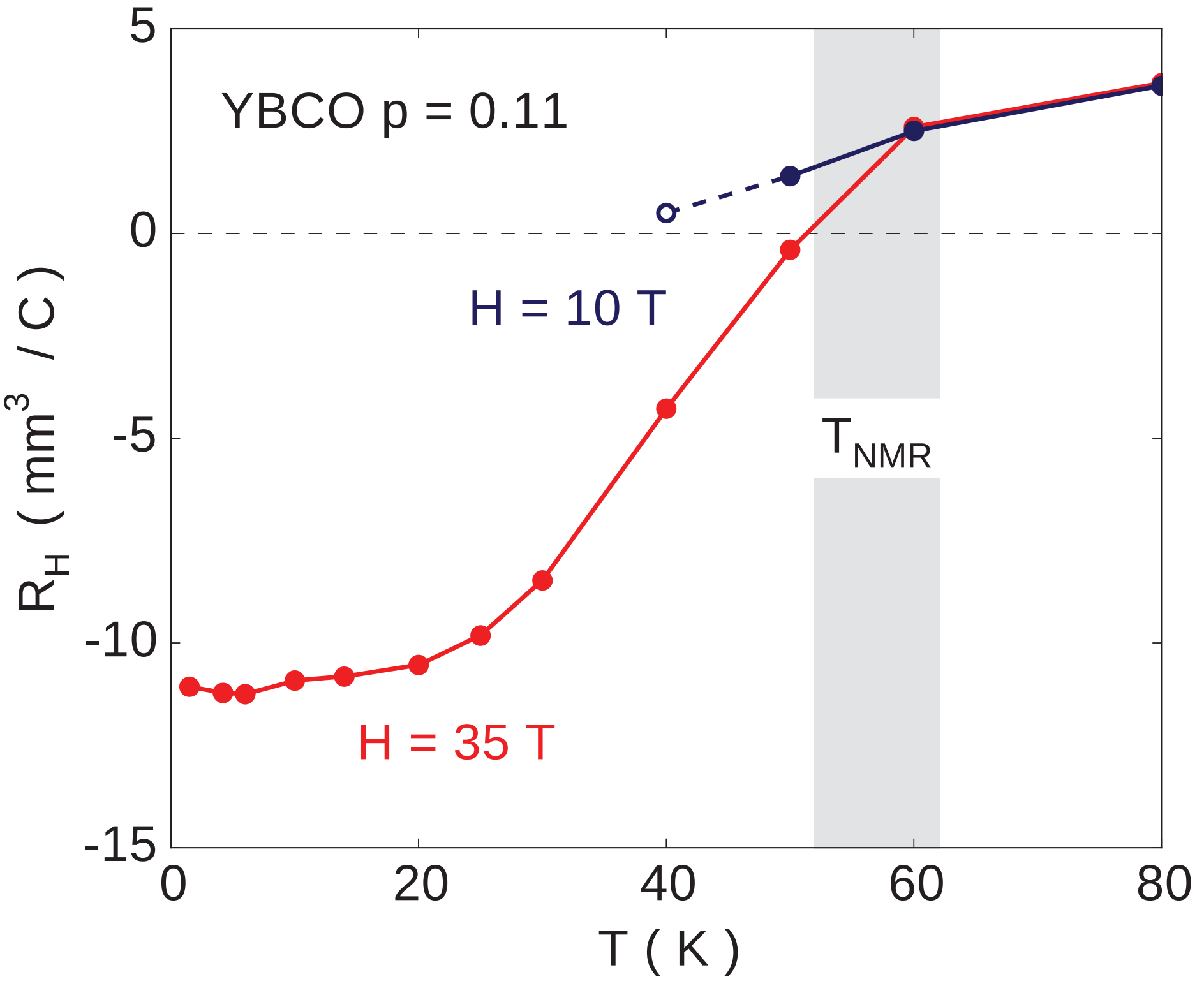

**Figure S6 | Electrical Hall coefficient of YBCO at *p* = 0.11.**

Temperature dependence of the Hall coefficient in YBCO at $p$ = 0.11 ($T_c$ = 61 K), plotted as $R_H$ vs $T$ for $H$ = 10 T < $H_{FSR}$ (blue circles) and $H$ = 35 T > $H_{FSR}$ (red circles). The open blue circle is below $T_c(H$ = 10 T) ~ 50 K. Above 60 K, both curves share a common slow decrease with cooling. We attribute this field-independent decrease to a FSR caused by the field-independent CDW modulations whose correlation length increases upon cooling[16,17], in a regime where the electronic mean free path is very short. Below 60 K, the two curves split off, with the 35 T curve dropping much more rapidly, consistent with the onset of CDW order detected by NMR above a threshold field $H_{NMR}$ = 10 T, below a transition temperature $T_{NMR}$ = 57 ± 5 K (ref. 9; grey band).